\documentclass[12pt]{iopart}

\usepackage{color}
\usepackage{cite}
\usepackage{amssymb}
\usepackage{graphicx}   
\usepackage{subfigure}
\usepackage{bm}

\begin{document}     
	\title[Graphene Pt Nanoparticle Multilayers]{Multilayer Structures of Graphene and Pt Nanoparticles - a Multiscale Computational Study}
	\author{Samaneh Nasiri$^1$, Christian Greff $^1$, Kai Wang$^2$, Mingjun Yang$^2$, Qianqian Li $^3$, Paolo Moretti $^1$, Michael Zaiser $^1$$^,$$^2$$^,$$^4$ }
	
	\address{$^1$WW8-Materials Simulation, Department of Materials Science, Friedrich-Alexander Universit\"at Erlangen-N\"urnberg, 90762 F\"urth, Germany} 
	\address{$^2$School of Materials Science and Engineering, Southwest Petroleum University, Chengdu, Sichuan, PR China}
	\address{$^3$Department of Aerospace Engineering, Imperial College, Prince Consort Road, London SW7 2BZ, UK}
	\address{$^4$School of Mechanics and Engineering, Southwest Jiaotong University, Chengdu, Sichuan, PR China}
	\ead{samaneh.nasiri@fau.de}

\begin{abstract}
We report results of a multiscale simulation study of multilayer structures consisting of graphene sheets with embedded Pt nanoparticles. Density functional theory is used to understand the energetics of Pt-graphene interfaces and provide reference data for the parameterization of a Pt-graphene interaction potential. Molecular dynamics simulations then provide the conformation and energetics of graphene sheets with embedded Pt nanoparticles of varying density, form and size. These results are interpreted using a continuum mechanical model of sheet deformation, and serve to parameterize a meso-scale Monte Carlo model to investigate the question under which conditions the free volume around the Pt nanoparticles forms a percolating cluster, such that the structures can be used in catalytic applications. We conclude with a discussion of potential applications of such multilayer structures. 
\end{abstract}

\section*{Keywords}
Graphene, Platinum, Catalyst, Multiscale Simulation, Nanoparticles.  

\maketitle

\section{Introduction}

It is well known that  \(sp^2\) bonded carbon nanoparticles (CNP), in particular graphene (GP) flakes and carbon nanotubes (CNT), possess excellent mechanical properties \cite{lee2008-science,zhao2002-PRB,nasiri2016-AIMSmaterialsscience}, high specific surface area \cite{zhang2013-Sci.Rep, stoller2008-Nanoletters}, good thermal properties \cite{balandin2011-Nat.Mater,pop2012-MRSbulletin} and high electrical conductivity\cite{neto2009-Rev.Mod.Phys.,marinho2012-PowderTechnol}. These outstanding properties make graphene a suitable candidate to act as catalyst support for metals and non-metals in electro-catalytic applications such as use in fuel cells and batteries\cite{antolini2012-ApplCatalB,brownson2011-J.PowerSources,choi2012-NanoEnergy,wang2013-SCR}. Since the discovery of graphene there has been a significant amount of research on using platinum decorated graphene sheets as new catalysts in low temperature fuel cells because of their high stability and electro-catalytic ability and more importantly their high catalyst loading capabilities \cite{li2009-Electrochem.Comm.,liu2010-J.PowerSources,samad2018-Int.J.Hydrog.Energy,seselj2015-Sciencebulletin,dong2010-Carbon,zhao2015-J.Mater.Chem.A,Ruguang2019-npj}. 

Stability of metal nanoclusters on carbon based catalysts is one major issue in catalytic applications. Due to the \(sp^2\) carbon bonds, pristine graphene is relatively inert. Metal nanoparticles adsorbed onto graphene are found to be fairly mobile on the surface, indicating weak bonding \cite{khomyakov2009-PRB, schneider2013-ChemPhysChem, ramos2013-Phys.Chem.Chem.Phys, gan2008-Small}. Using dispersion-corrected DFT methods, Schneider et al.\cite{schneider2013-ChemPhysChem} showed that for nanoclusters larger than about 1 nm, the interaction between platinum and graphene is purely of van der Waals type. As the weak bonding affects the mechanical stability of the catalyst, methods have been sought to cause stronger attachment of the Pt clusters to the graphene sheets. Such methods include heat treatment \cite{yoo2009-Nanoletters}, defect engineering \cite{fampiou2012-J.Phys.Chem, schneider2013-ChemPhysChem, cheng2014-Acta} as well as chemical surface modification of graphene \cite{jafri2015-Int.J.Hydrog.Energy,mazanek2018-Chemistry-A,wang2013-ACS}.  

An alternative approach towards enhancing the attachment between Pt nanoparticles and graphene support is to create sandwich structures in which Pt nanoparticles are anchored between the two adjacent graphene sheets. Such structures were found to possess higher activity for methanol electrooxidation and a higher stability than that of Pt-graphene\cite{zhao2015-J.Mater.Chem.A}. Somewhat paradoxically, while mutually adhering graphene sheets have been argued to hold Pt nanoparticles in place, it has also been argued that discrete metal nanoclusters on graphene sheets can act as geometrical nano-spacers to reduce or prevent graphene agglomeration \cite{chang2014-Nanoscale,pasricha2009-Small,cho2018-Journal.Chem.Eng,buglione2012-ChemPlusChem}. For instance, it has been demonstrated that decoration of exfoliated graphene sheets with Pt nanoclusters can prevent face-to-face aggregation of the sheets \cite{si2008-Chem.ofMat}. These apparently contradictory applications -- Pt nanoclusters reduce graphene-graphene adhesion and prevent graphene agglomeration, mutually adhering graphene sheets hold Pt nanoclusters in place -- indicate a need for a systematic investigation. The aim is to establish, for Pt nanoclusters of different sizes and densities embedded between graphene sheets, the mechanical stability and energetic properties of the resulting multilayer structures in order to decide whether they are promising for use in catalytic applications. 

In the present work, we report results of a multiscale simulation study of bilayers of graphene with embedded Pt nanoparticles. We first perform density functional theory (DFT) calculations to establish the energetics of Pt-graphene interfaces and to obtain reference data for parameterizing interaction potentials for the Van der Waals - like Pt-graphene interactions. We then use these potentials in a molecular mechanics (MM) and molecular dynamics (MD) study to establish sheet conformations of graphene bilayers with embedded Pt clusters of varying size. These simulations are analyzed using a continuum model to establish analytical relations between the density and the size of embedded Pt nanoparticles and the effective adhesion properties of the bilayers. This analysis is complemented by a Monte-Carlo model to investigate under which conditions the free volume surrounding the nanoparticles in such structures forms a percolating cluster as required for their catalytic use. 

\section{Atomic-scale simulations}
\label{atomistic}

On the atomic scale, we first perform DFT calculations of planar graphene-Pt interfaces in order to gain data for parameterizing empirical potentials that are then used, in larger-scale MD simulations, to describe the Van der Waals-like interaction between graphene sheets and Pt nanoparticles. 

\subsection{Density-functional calculations: Methods and results}

\begin{figure*}
	\centering
	\resizebox{0.8\textwidth}{!}{
	\includegraphics{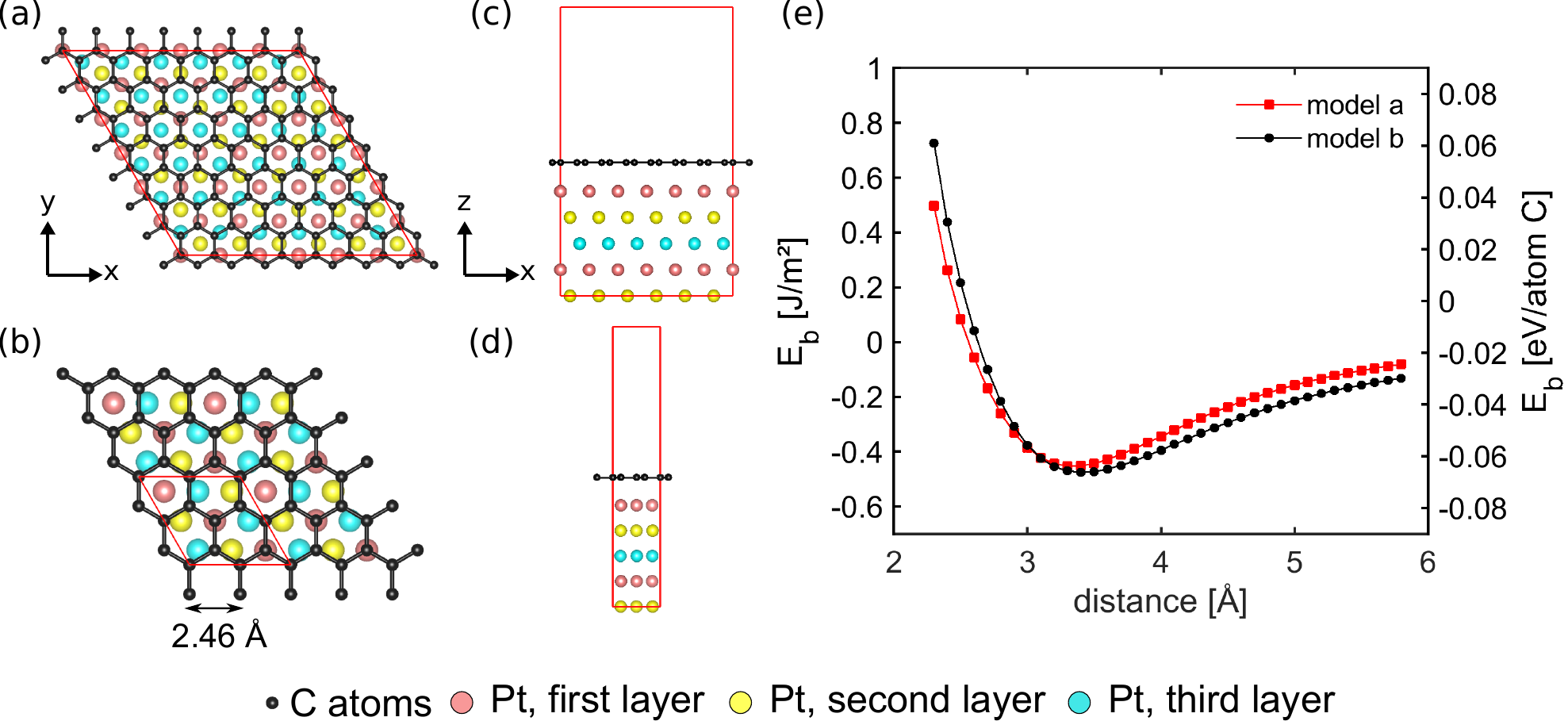}}
	\caption{Top views ((a) and (b)) and side views ((c) and (d)) of the geometry for model a and model b respectively. (e) Pt-graphene binding energy for both models obtained by DFT calculations.}
	\label{fig:1}       
\end{figure*}

In order to calculate the Pt-graphene interaction using DFT we consider five-layer Pt(111) slabs adhering to monolayer graphene. Both Pt slabs and graphene sheet are oriented parallel to the $xy$ plane of a Cartesian coordinate systems. Periodic boundary conditions are used in all three spatial directions with a 10 \AA{} vacuum gap between the periodic replicas in the $z$ direction. Two variants of this structure (graphene on Pt (111) surface) are constructed as shown in Figure \ref{fig:1}; for model a, a larger supercell consisting of p \((7 \times 7)\) graphene and p \((6 \times 6)\) Pt slabs were used in the calculations to mimic an incoherent interface. The structure for model b consists of p \((2 \times 2)\) graphene and three atoms per Pt layer and represents a coherent interface as considered in previous studies \cite{gong2010-J.Appl.Phys.,khomyakov2009-PRB}. For both models, the Pt lattice parameters were set to experimental values for bulk Pt. 

All the calculations were carried out by the Quantum Espresso package version 6.1 \cite{giannozzi2009-J.Phys.Condens.Matter}, employing PBE-based projected augmented wave(PAW) potentials \cite{blochl1994-PRB}. Previous studies have shown that the optB88-vdw \cite{becke1988-PRA} exchange-correlation functional gives a reasonably accurate prediction of both interlayer distance and binding energy for graphite  \cite{graziano2012-J.Phys.Condens.Matter} , so here we choose the optB88-vdw functional in all our DFT calculations. In our calculations, the kinetic energy cutoffs of 40 Ry and 450 Ry were used for wave function and charge density calculations respectively, and a Methfessel-Paxton smearing of 0.01 Ry was used for the electronic convergence. The convergence for self-consistency calculations was less than 0.0001 Ry. In the calculations of model a, only the gamma k-point was used due to the large system we chose, while an \(8\times8\times1 \) k-points grid was used in the calculations of model b.

Since our aim is to obtain data for parameterizing an interaction potential, rather than performing an accurate analysis of the interface structure and energy using DFT, we do not carry out any structural relaxation. Instead we simply separate the graphene sheet rigidly from the Pt(111) slab without geometry relaxation of either the Pt slab or the graphene. We evaluate the binding energy per carbon atom as
\begin{eqnarray}
E_{\rm b} = [E_{\rm tot} - (E_{\rm pt} +E_{\rm gr})]/n \label{eq1} 
\end{eqnarray}

\(E_{\rm tot}\) is the total energy of the graphene-Pt(111) system, \(E_{\rm pt}\) represents the energy of the five-layer Pt(111) slab, and \(E_{\rm gr}\) represents the energy of the free-standing graphene sheet. Results are depicted in Figure \ref{fig:1} for both models.

\begin{figure}
	\centering
	\includegraphics[width=0.4\textwidth]{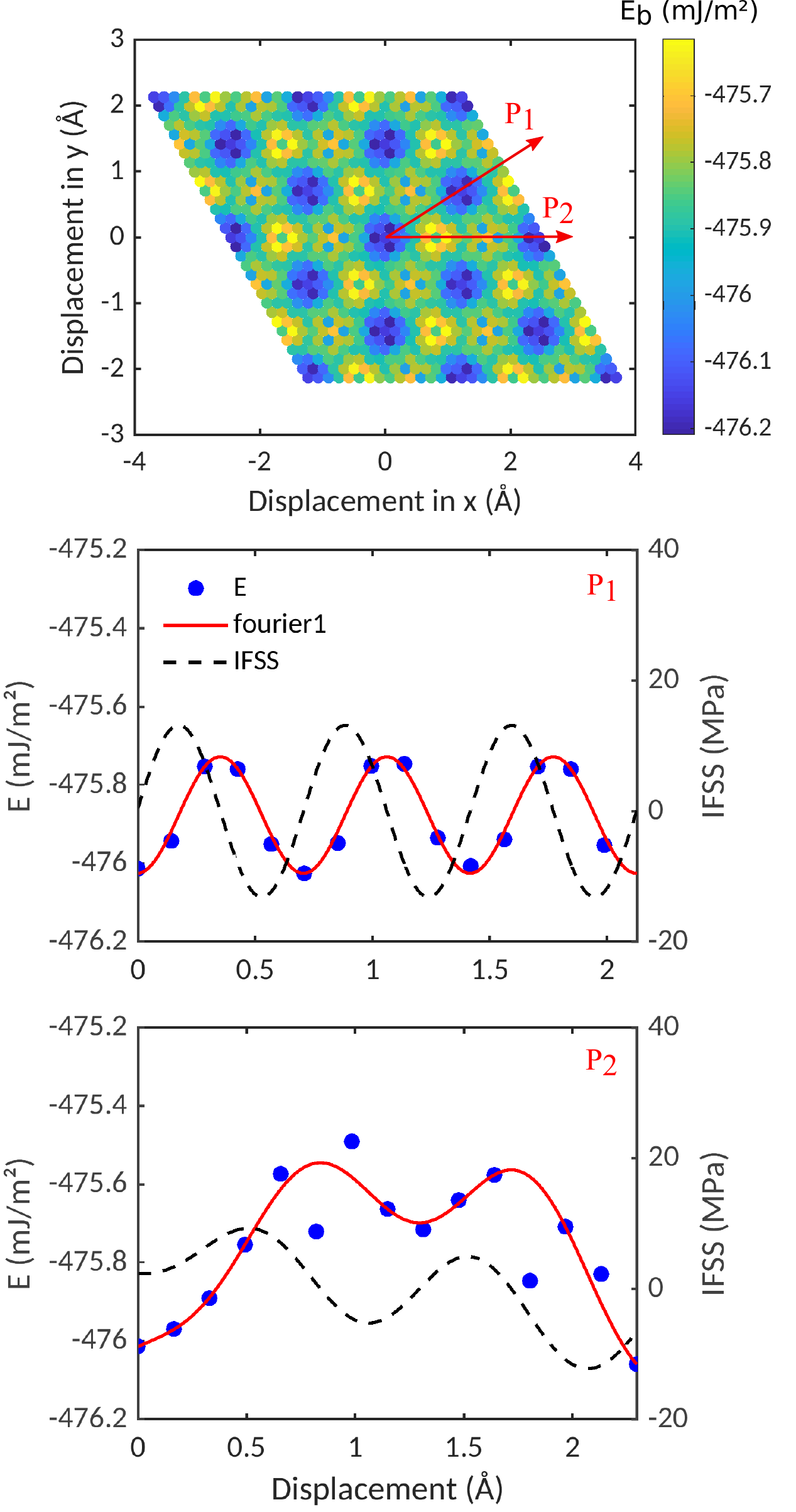}
	\caption{Top: Interface energy surface obtained by sliding graphene rigidly on top of a Pt slab, using model (b) and keeping the distance at the position of lowest energy; bottom: Interface energy profiles and interface shear stress profiles taken along the displacement paths \(\rm P_{\rm 1}\) and \(\rm P_{\rm 2}\).}
	\label{fig:1a}      
\end{figure}

To obtain an estimate of the stress needed to shear the graphene-Pt(111) interface, we also evaluate the 'interface energy surface' for model (b) which is obtained by sliding the graphene sheet rigidly on the Pt surface. Results are shown in Figure \ref{fig:1a} in conjunction with energy profiles taken along three typical paths. It can be seen that the interface energy variations are quite small. Accordingly, the interface shear stresses (IFSS) -- derivatives of the energy profiles -- required to slide the graphene over the Pt surface are comparatively low, of the order of 20 MPa or less. The same is expected to hold for the typical shear stresses required to slide a Pt cluster on a graphene sheet. 

\subsection{Molecular simulation: Validation of interaction potentials}

We use the LAMMPS simulation package \cite{plimpton1995-J.Comput.Phys.} to perform molecular mechanics (MM) and molecular dynamics (MD) simulations of binary systems consisting of platinum nanoparticles which are anchored between two layers of graphene sheets. For describing such a system we need to define Pt-Pt, Pt-C and C-C interaction potentials, where the latter have to account for both short and long range interactions between carbon atoms within the graphene sheets and between both sheets.

In a  first step, we examine different sets of interaction potentials to establish which set of potentials is most appropriate for the problem at hand. The first potential we examined is the reactive force field (ReaxFF) which has been developed by Sanz-Navarro et al. \cite{sanz2008-J.Phys.Chem.A,aktulga2012-parallel}. ReaxFF is based on a bond-order model in conjunction with a charge-equilibration scheme\cite{rappe1991-J.Phys.Chem.A}. This potential is in principle able to describe all the above mentioned interactions in the Pt-C system, however, it is optimized to correctly represent reactive processes rather than mechanical properties. The second potential we considered is a Brenner force field (BrennerFF) parametrized by Albe et al. \cite{albe2002-PRB} to describe short-range interatomic interactions in the Pt-C systems, with a Lennard-Jones potential added to mimic the weak van der Waals interaction between the carbon atoms of the two graphene sheets \cite{brenner2002-J.Phys.Condens.Matter}. A further possibility is to combine potentials of different type that are separately optimized to describe the energetics and mechanical properties of covalently bonded carbon, and of platinum. A similar approach was used by \cite{kim2013-Nature , nasiri2019-EPJ} for simulating Ni-graphene interactions: These authors use a standard AIREBO potential \cite{brenner2002-J.Phys.Condens.Matter,stuart2000-J.Chem.Phys.} for describing covalent C-C interactions in conjunction with an EAM potential for Ni, and a Lennard-Jones potential for Ni-carbon interactions at the Ni-graphene interface.  We emphasize that such addition of structurally dissimilar potentials is, in general, problematic. However, in the present context the approach may be feasible because the components of the system (metal, graphene) maintain their structural integrity, and the van der Waals type interaction between these components can be considered as additive to their internal (metallic, covalent) interactions. Accordingly we consider a Hybrid combination comprising the standard AIREBO potential \cite{brenner2002-J.Phys.Condens.Matter,stuart2000-J.Chem.Phys.} for covalent C-C interactions and the EAM potential of Zhou and coworkers \cite{zhou2004-PRB} for Pt-Pt interactions. Pt-C interactions at the interface are described using either a LJ potential taken from the literature \cite{huang2003-SurSc}; this combination is denoted as HybridFF-LJ. Alternatively we consider a Morse potential of the form $E_{\rm M} = D (\exp[-2\alpha (r - r_{\rm 0})] - 2 \exp[-\alpha (r - r_{\rm 0})])$ where $r$ is the Pt-C distance,  $r_0$  an equilibrium bond distance,  $D$ is the well depth, and  $\alpha$ controls the stiffness of the potential (the smaller $\alpha$ is, the smaller the attractive/repulsive forces). The combination of AIREBO (C-C) and EAM (Pt-Pt) with this Morse potential, whose parameters we determine by matching with our own DFT calculations, is denoted as HybridFF-M. 

The performances of the different potentials or combinations of potentials are compared in view of two main criteria. First, we ask how well the potentials reproduce theoretical predictions for the mechanical properties of the system components. Second, we study how well they describe the adhesion between Pt and graphene. Regarding mechanical properties,  the data in Table \ref{Pt} show that all potentials correctly represent the lattice constant and cohesive energy of Pt, however, the ReaxFF produces isotropic elastic behavior for Pt, completely disregarding the significant cubic anisotropy of the material. The other potentials yield acceptable results regarding the elastic properties of Pt.  Lattice properties of monolayer graphene as well as the interlayer distance \(d_{\rm 0}\) and interlayer adhesion energy $\gamma_{\rm GG}$ of bilayer graphene are reported in Table \ref{C}.

\begin{table}
	\centering
	\caption{Lattice properties of platinum.}
	\label{Pt}
	\begin{tabular}{lllll}
		\hline\noalign{\smallskip}
		&EAM & ReaxFF & BrennerFF & Experiment\cite{collard1992-acta,rassoulinejad2018-nature} \\
		\noalign{\smallskip}\hline\noalign{\smallskip}
		\(a_{\rm 0}\)(\AA{}) &3.92&3.94&3.92&3.91\\
		\(E_{\rm 0}\)(\rm{eV})&-5.77&-5.72&-5.77&-5.77\\
		$C_{11}$(GPa) &357.11&332.02&351.91&373.42\\
		$C_{12}$(GPa)&260.13&194.63&248.45&241.74\\
		$C_{44}$(GPa)&77.97&194.63&89.94&77.65\\
	\end{tabular}	
\end{table} 

\begin{table}
	\centering
	\caption{Lattice properties of graphene.}
	\label{C}
	\begin{tabular}{lllll}
		\hline\noalign{\smallskip}
		 &ReaxFF & BrennerFF & AIREBO & DFT\cite{spanu2009-PRL}\\
		\noalign{\smallskip}\hline\noalign{\smallskip}
		\(a_{\rm 0}\)(\AA{}) &2.46&2.46&2.46&2.45\\
		\(E_{\rm 0}\)(\rm{eV})&-7.46&-7.35&-7.39&-7.46\\
		\(d_{\rm 0}\)(\AA{}) &3.29&--&3.41&3.35\\
		\(\gamma_{\rm GG}\)$(\rm J/m^2)$&-0.42&--&-0.29& -0.25 -- -0.36\\
	\end{tabular}
\end{table} 

Concerning the binding between Pt and graphene, we use our DFT results for the interaction between Pt(111) and graphene as a reference. We consider the same configuration as in the DFT simulations, i.e., we separate the graphene sheet rigidly from the Pt(111) surface, which allows us to directly compare the MD and DFT energies. First, we observe that the BrennerFF completely fails to reproduce the interfacial bonding between graphene and Pt(111) -- the interaction reduces to a short-range repulsion and the binding energy is thus zero -- which renders this potential unsuitable. The ReaxFF significantly over-estimates the bonding between Pt and graphene while it under-estimates the interfacial separation. Therefore, and also in view of its poor performance in modeling the elastic properties of Pt, this potential is also ruled out, which leaves us with the two hybrid combinations of potentials (HybridFF-LJ and HybridFF-M). 

\begin{figure}
	\centering
	\includegraphics[width=0.5\textwidth]{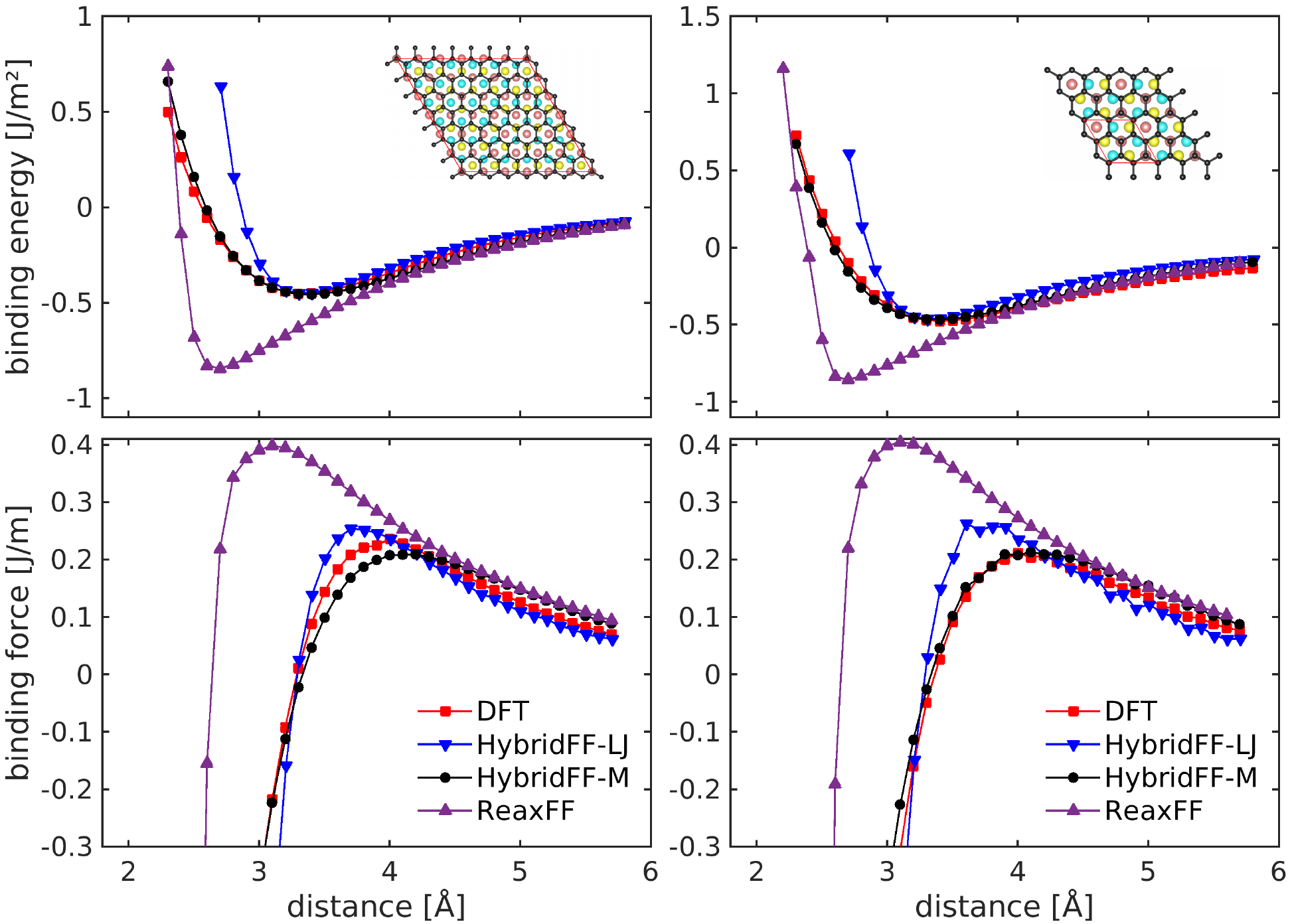}
	\caption{Binding force and energy for interface model a (left) and model b (right); comparison of DFT data with MD results for different interaction potentials. }
	\label{fig:2}       
\end{figure}

The two hybrid potentials  offer the advantage that the parameters of the LJ or Morse potential, which describes Pt-C interactions at the interface between graphene sheets and Pt, can be fitted to correctly reproduce the interface separation and binding energy. At variance with previous studies \cite{huang2003-SurSc}, we find that the LJ potential has clear deficiencies as it is too stiff and thus significantly over-estimates the forces acting across the Pt-graphene interface (see Figure \ref{fig:2} (c,d)). The Morse potential is the only one that can be fitted very well to the DFT data because the well depth, well width and equilibrium distance parameters can be optimized to reproduce the quantum mechanical interaction energy curve from the density functional calculations. The resulting parameters for the Pt-C Morse potential are $D$ = 0.0071 eV, $r_{\rm 0}$ = 4.18 \AA{} and $\alpha$ = 1.05 $\mathrm{\AA}^{\rm -1}$.

We therefore use in the following the HybridFF-M combination of potentials: The AIREBO potential \cite{brenner2002-J.Phys.Condens.Matter,stuart2000-J.Chem.Phys.} for covalent C-C interactions, the EAM potential of Zhou and coworkers \cite{zhou2004-PRB} for Pt-Pt, and the DFT parameterized Morse potential for C-Pt interactions. Van der Waals interactions between different graphene sheets are described by a standard LJ potential. 

In nanoscale clusters, the configuration of Pt atoms differs from the planar slab used in our reference DFT calculation because many atoms have reduced coordination number and/or local environments of reduced symmetry. To verify that the HybridFF-M combination of potentials correctly describes the adhesion of small Pt clusters to graphene, we consider the extreme case of a very small $\mathrm{Pt}_{\rm 13}$ nanocluster adhering to a pristine graphene sheet. DFT calculations for this system were performed by Fampiou and Ramasubramanian \cite{fampiou2012-J.Phys.Chem}
who report an adhesion energy of -0.84 eV for a minimum energy configuration of non icosahedral symmetry. Repeating the calculation with the same structure file but using the HybridFF-M potential for energy evaluation gives an adhesion energy of -0.94 eV, i.e., the deviation from the DFT result is only 11\%. (We note that calculations using HybridFF-LJ, Tersoff and ReaxFF yield discrepancies of over 300 \% of the DFT adhesion energy.) This gives us confidence that the HybridFF-M combination used in the following MM and MD simulations adequately captures the adhesion of larger Pt clusters on graphene. 

\begin{figure}
	\resizebox{0.5\textwidth}{!}{us 
		\includegraphics{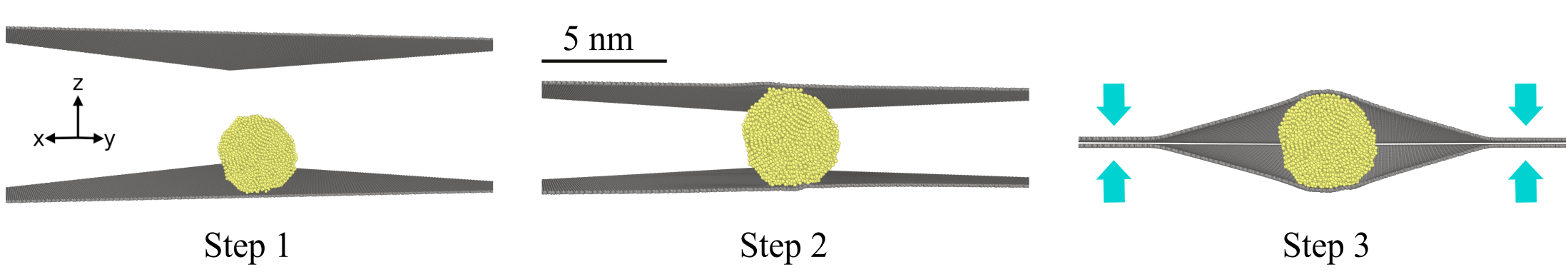}}
	\centering
	\caption{Modeling the conformation of graphene bilayer around an annealed Pt particle using MM simulation.}
	\label{fig:3}       
\end{figure} 

\subsection{MM and MD simulations of Pt-Graphene structures: Set-up and results}

For simulating the energy and geometrical properties of graphene-encapsulated Pt nanoparticles, we consider two types of particles: On the one hand, we use Pt particles constructed with cylindrical geometry (radius $R_{\rm p}$, particle height $2h$) to produce a prescribed aspect ratio of $ q_{\rm p}=h/R_{\rm p}$. On the other hand, we consider annealed particles where an initially cylindrical particle is, before encapsulation, heated above the melting point of Pt and slowly cooled down to near 0\,{}K using a Nos\'e{}-Hoover-thermostat in order to minimize the particle energy. Such annealed particles generally exhibit a spheroidal particle shape. In order to encapsulate the particles, we start from two graphene sheets distanced significantly beyond the cut-off radius in order to avoid van der Waals interaction between the layers. One platinum particle is positioned on the lower graphene sheet within the cut-off radius of the Pt-C interaction (Figure \ref{fig:3}: step 1). To prevent edge effects, periodic boundary conditions are used in all directions. To form a system where the platinum particle is acting as a spacer between the two graphene sheets, the upper graphene sheet is rigidly moved right above the platinum particle (Figure \ref{fig:3}: step 2). Then, the outer areas of the two sheets are brought into contact with each other with 3.4\,$\mathrm{\AA}$ distance between them. A minimization step brings the system into mechanical equilibrium, where the platinum particle functions as a mechanical obstacle that prevents the two graphene sheets from adhering to each other over a 'detachment area' surrounding the particle (Figure \ref{fig:3}: step 3). In all simulations, we determine the total energy of the relaxed structure as well as the elastic energy of the graphene sheet and the energies of graphene-graphene, Pt-Pt, and graphene-Pt interaction. In addition, we determine the detachment area $A_{\rm d}$ as the area over which the spacing between the graphene layers exceeds the value for pure bilayer graphene by more than 1 \AA, and for single encapsulated particles we define the detachment radius via $A_{\rm d} =: \pi R_{\rm d}^2$. 
\begin{figure*}
	\centering
	\includegraphics[width=0.8\textwidth]{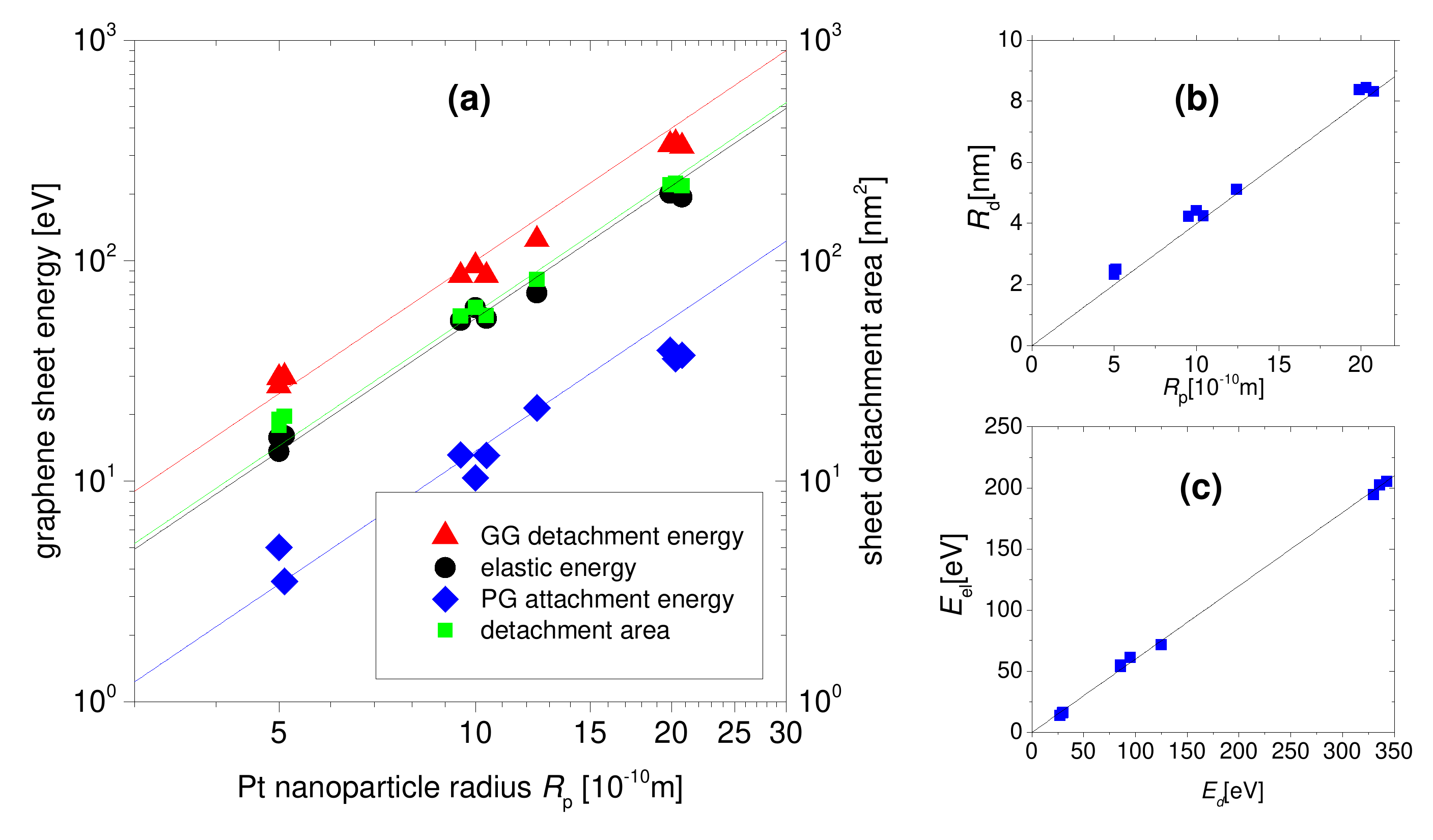}
	\caption{(a) Detachment area, energy of GG detachment, energy of PtG attachment, and graphene elastic energy as functions of Pt nanoparticle radius; data points: numerical values obtained from our MM simulations; solid red line: $E_{\rm d}$ after Eq. (6) with $q_{\rm d} = 4.0$, 
solid black line: $E_{\rm el}$ after Eq. (7) with $q_{\rm d}= 4.0$; solid blue line: $E_{\rm a}$  after Eq. (8) with $q_{\rm c} = 0.87$; (b) Detachment radius as function of particle radius, solid line: $R_{\rm d} =  q_{\rm d} R_{\rm p}$ with $q_{\rm d}=4.0$; (c) Elastic energy as a function of detachment energy, solid line: Eq. (7) with $q_{\rm d}= 4.0$.}
	\label{fig:3a}       
\end{figure*}   
MM simulations of single embedded Pt nanoparticles of varying radius but constant aspect ratio indicate that the detachment radius surrounding a particle is proportional to the particle radius, i.e. $R_{\rm d} =  4.2 R_{\rm p}$ (Figure \ref{fig:3a}b). The energy of the graphene bilayer is raised by embedding a Pt nanoparticle. This energy increase ('GG excess energy') is due to the energy of adhesion which must be provided in order to detach the two graphene sheets over the detachment area (detachment energy: $E_{\rm d}$), diminished by the adhesion between the two graphene sheets and the nanoparticle (attachment energy: $E_{\rm a}$), and due to the elastic distortion of the sheets (elastic energy: $E_{\rm el}$). All three energy contributions, which we evaluate from the MM/MD simulations, are found to increase in mutual proportion as the Pt particle radius increases. A simplified continuum model formulated in section 3 allows to understand this behavior.

\begin{figure}
	\centering
	\includegraphics[width=0.48\textwidth]{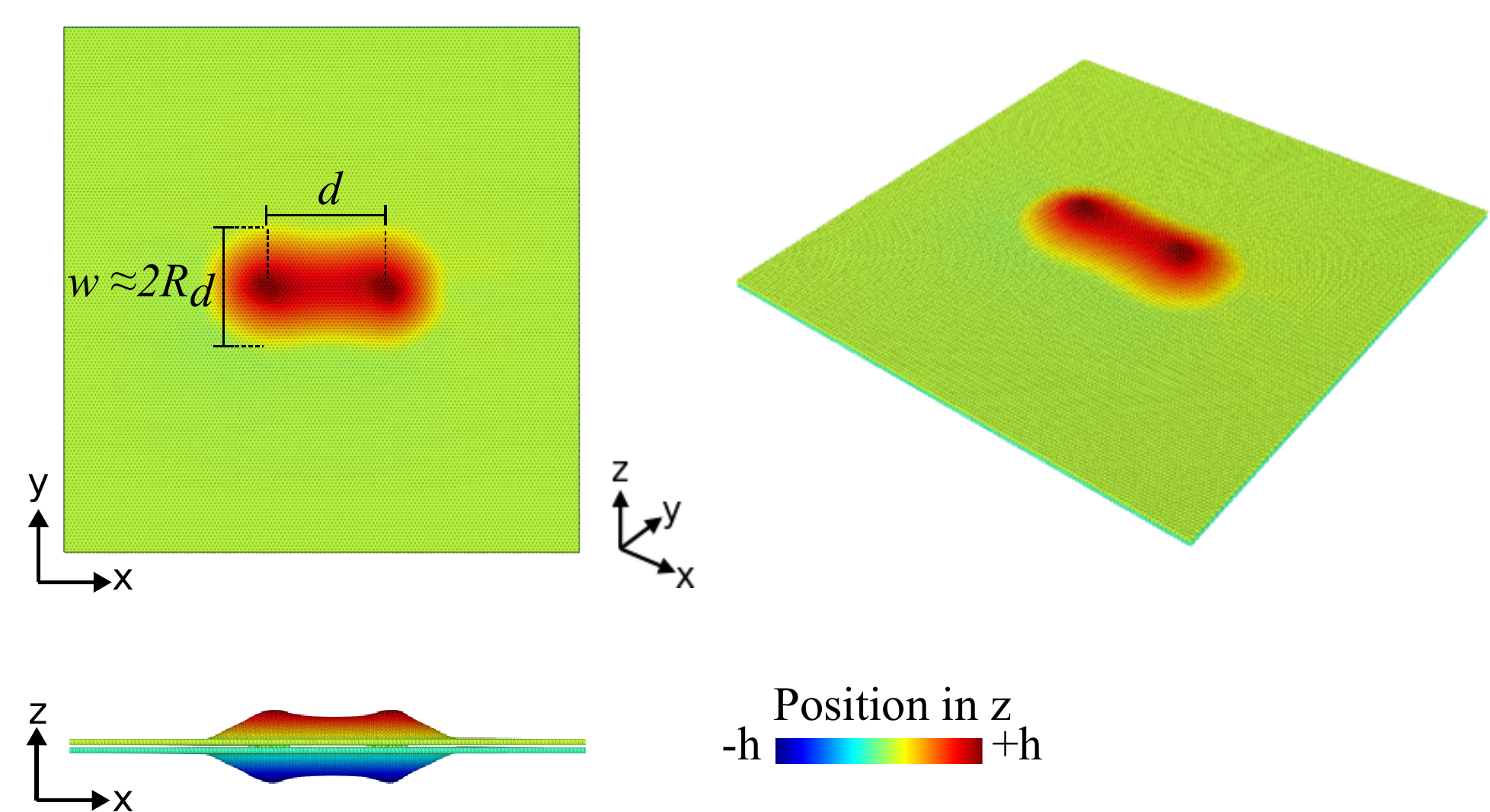}
	\caption{Molecular mechanics simulation of a two-nanoparticle system; the coordinates of the Pt atoms were kept fixed during relaxation of the graphene sheets.}
	\label{fig:3b}       
\end{figure}  

\begin{figure}
	\centering
	\includegraphics[width=0.45\textwidth]{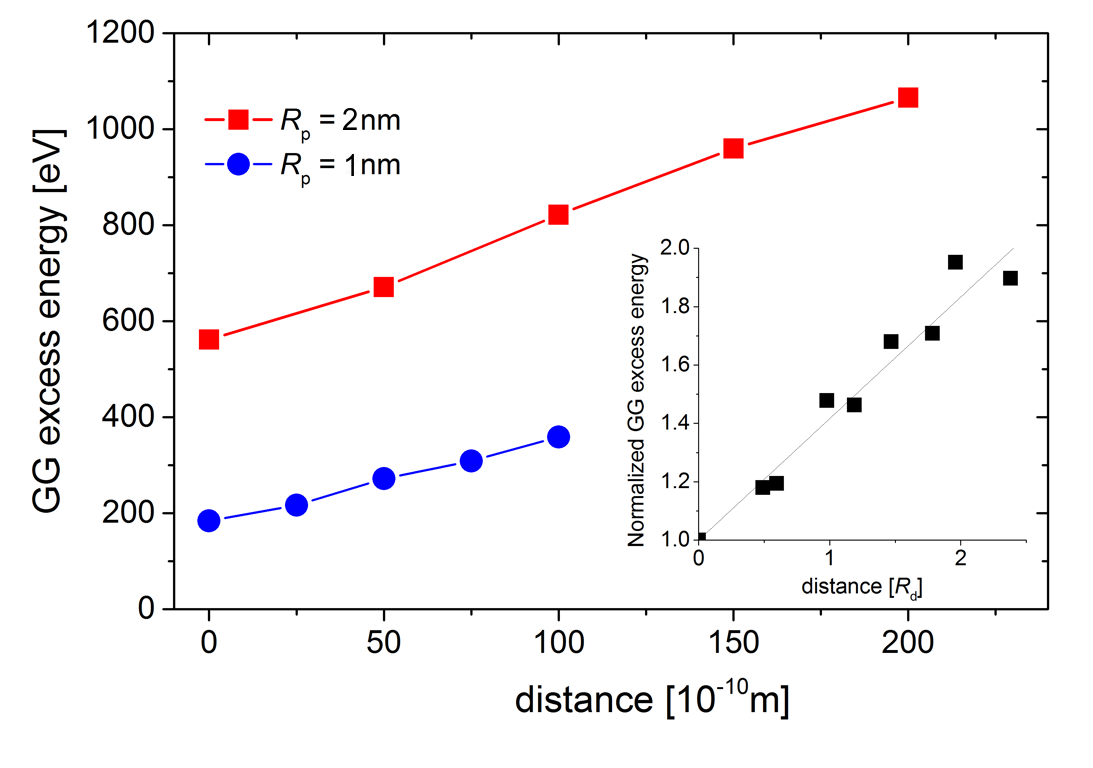}
	\caption{GG excess energy of two-nanoparticle systems as function of nanoparticle separation distance; red symbols: $R_{\rm p}=2$nm, 
	blue symbols: $R_{\rm p}=1$nm; inset: normalized excess energy, measured in units of the GG excess energy around a single particle, vs. inter-particle distance in units of the single-particle detachment radius.}
	\label{fig:3c}     
\end{figure}

In order to establish the dependence of the GG excess energy on the distance between Pt particles, we perform calculations on nanoparticle pairs as shown in Figure \ref{fig:3b}. To control the inter-particle separation distance $d$, in these MM simulations we keep the coordinates of the Pt atoms fixed during the final relaxation step. If the detachment areas surrounding two Pt nanoparticles overlap, this leads to a reduction in GG excess energy since the overall detachment area is reduced and also the elastic deformation of the graphene sheet decreases. The energy reduction becomes more pronounced as the distance between the two particles decreases, which leads to an attractive configurational force acting on the particles. Curves showing the GG excess energy of two-nanoparticle systems ($E^{\rm 2p}$) with different nanoparticle size vs the nanoparticle separation $d$ are shown in Figure \ref{fig:3c}. A data collapse can be obtained when the GG excess energy is scaled by its value for a single particle ($E^{\rm 1p}$), and the inter-particle separation is measured in units of the detachment radius $R_{\rm d}$ of the graphene sheet around that particle. The energy vs. distance curve can be well approximated by a linear dependency, $E^{\rm 2p} = E^{\rm 1p}(1+d/d_{\rm c})$ where $d \approx 2.4 R_{\rm d}$ is the critical distance where the particles first interact with each other. Above that distance, we are dealing with two separate 
single particles (see Figure \ref{fig:3c} inset). 

\section{Continuum mechanical model}
\label{CM}
In the following we use the results of our MM/MD simulations to parameterize a simplified continuum mechanical model which transfers the atomic
interactions into continuum-level concepts such as elastic energy, work of adhesion, and interfacial shear stress acting between continuous bodies. Such a model is necessarily inadequate for very small Pt clusters (an extreme case being $\mathrm{Pt}_{\rm 13}$) where atomic structure is crucial. However, we expect the continuum model to become more accurate as we move to larger and larger Pt nanoparticles. In fact,  the scaling relations between energies and geometrical parameters we expect from the continuum model extend down to clusters with a diameter as low as 1nm. This is demonstrated in Figure 5 by the comparison of the discrete data points (simulation data) and straight lines (continuum mechanical scaling relations). A continuum mechanical treatment is useful for several reasons: (i) through simplified calculations it  provides analytical estimates of energies and interactions in the system at hand; (ii) by their nature, continuum mechanical models are scalable and can be applied to larger nanoparticles without adding additional computational cost; (iii) numerical implementation of such models (not studied here) can be easily generalized to more complex multilayer architectures.

\subsection{ Model of a single embedded particle}

\begin{figure}
	\centering
	\includegraphics[width=0.45\textwidth]{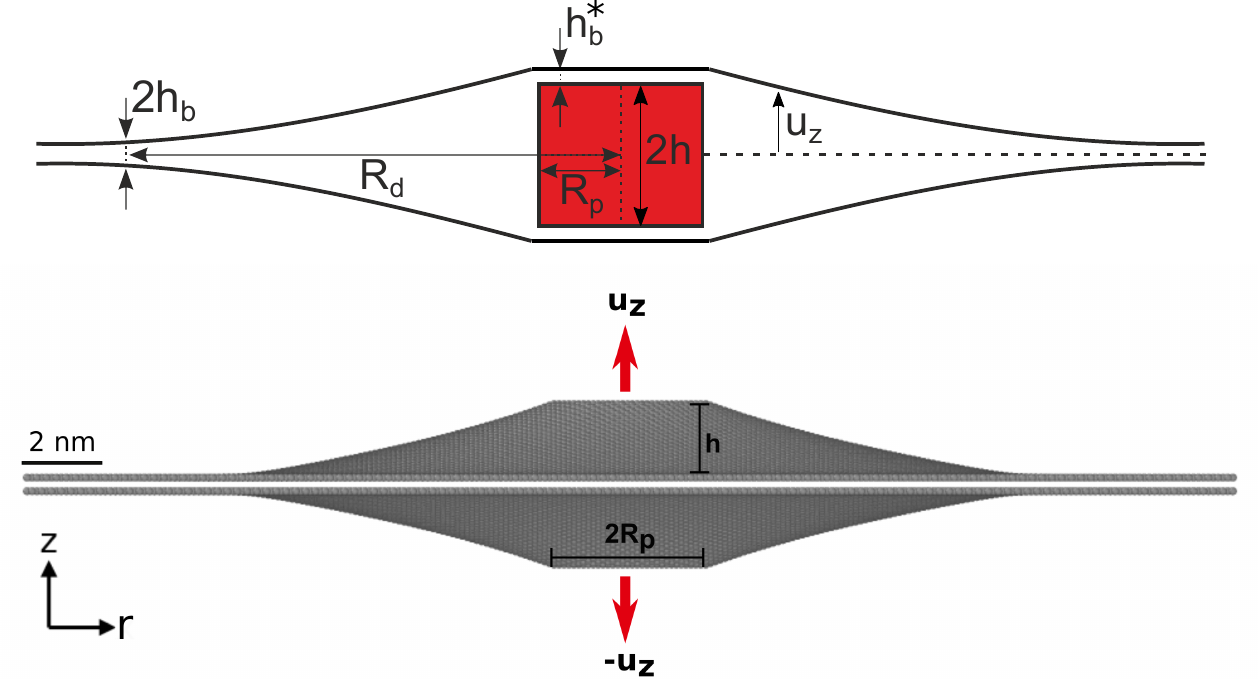}
	\caption{Schematic of the continuum mechanical model (top) and reference MD model (indentation model) (bottom).}
	\label{fig:4}       
\end{figure}

To obtain semi-analytical relations between the size of the Pt nanoparticles encapsulated between the graphene sheets, the detachment between the two sheets, and the energy of the system, we use a simplified continuum model as illustrated in Figure \ref{fig:4}, top.  The particle is idealized as a cylindrical body with a radius of $R_{\rm p}$ and a height of $2h$, with its axis perpendicular to the adhering graphene sheets. Accordingly, a cylindrical coordinate system is used with its origin in the particle center. The $z$-axis runs along the particle axis, and the radial coordinate $r$ and angular coordinate $\theta$ denote the position parallel to the the plane $z=0$, which corresponds to the plane of adhesion between the two graphene sheets. The deformation of the graphene sheets is described by the displacements $u_r, u_{\theta{}},u_z$. Pt-graphene binding length ($h_{\rm b}^*$) and graphene-graphene inter-layer distance ($2h_{\rm b}$) are assumed to be equal. 
A matching atomistic model is provided by a graphene sheet indented by a cylindrical indenter of radius $R_{\rm p}$. In this model all atoms within $R_{\rm p}$ from the center of the bilayer sheet are displaced rigidly in the respective up- and downward direction by a distance $h$ while the rest of the sheet is allowed to relax (Figure \ref{fig:4}, bottom). 

The graphene sheets are modeled as 2D elastic membranes with a bending stiffness that is negligible compared to their in-plane stiffness. The latter is characterized by the planar elastic modulus $E_{\rm 2D}$ and Poisson number $\nu$. The radius of the detached area surrounding the particles is denoted as $R_{\rm d} > R_{\rm p}$. Because of the mirror symmetry of the problem with respect to the plane $z=0$, it is sufficient to consider a single membrane deformed by the half particle, hence, the problem is similar to the problem of  indentation of a 3D membrane of finite thickness as described by Begley \cite{begley2004-JMPS}, whose treatment we adapt to a 2D membrane sheet. We thus consider a membrane that is displaced, along the radius $r=R_{\rm p}$, by $u_z = h$ whereas for $r \ge R_{\rm d}$ the displacement is fixed at $u_z = 0$. A solution of the elastic problem with these boundary conditions is sought for $R_{\rm p} \le r \le R_{\rm d}$. The approximate solution of this elastic boundary value problem is discussed in supplementary information. Here we only summarize the main results:

\begin{itemize}
\item 
The displacement profile of the membrane is approximately given by
\begin{equation}
u_z(r) = h\frac{q_{\rm d}^{a}-(r/R_{\rm p})^{a}}{q_{\rm d}^{a}-1}\quad, q_{\rm d}=\frac{R_{\rm d}}{R_{\rm p}},\quad a \approx 2/3
\end{equation}
\item
The elastic energy of the system is given by
\begin{equation}
 E_{\rm el} = \pi R_{\rm p}^2  \frac{2E_{2D}}{27} \frac{q_{\rm p}^4}{(q_{\rm d}^{2/3}-1)^3}
 \end{equation}
\end{itemize}
The parameter $q_{\rm p}$ is the aspect ratio of the sandwiched particle. To determine the parameter $q_{\rm d}$ and hence the radius of the detached area surrounding the particle, we use a virtual work argument. The virtual work incurred upon expanding the detached area consists of the work released by the change in elastic energy and the work of adhesion that must be expended to detach the adhering graphene sheets. This detachment energy is simply given by ${\rm d}E_{\rm d} = 2\pi (\gamma_{\rm GG}/2) R_{\rm d} {\rm d} R_{\rm d}$ where $\gamma_{\rm GG}$ is the adhesion energy per unit area of the graphene bilayer, and $\gamma_{\rm GG}/2$ the adhesion energy per sheet. We thus set ${\rm d} W = {\rm d}E_{\rm el} + {\rm d}E_{\rm d} = 0$ which leads to the equation
\begin{equation}
\frac{{\rm d} E_{\rm el}}{dR_{\rm d}} +  \pi \gamma_{\rm GG} R_{\rm d} =0
\end{equation}
From this virtual work balance we derive an implicit relation connecting $q_{\rm p}$ and $q_{\rm d}$:
\begin{equation}
\frac{q_{\rm p}}{q_{\rm d}^{1/3}(q_{\rm d}^{2/3}-1)} = \left(\frac{27 \gamma_{\rm GG}}{4E_{2D}}\right)^{1/4} = 0.283
\end{equation}
where we used the adhesion energy $\gamma_{\rm GG} = 0.288$ J/m$^2$ for bilayer graphene and a planar elastic modulus of $E_{2D}=300$ N/m. Taking the detachment ratio for a given particle aspect ratio from this relation, we can write the detachment energy as a function of particle radius
\begin{equation}
E_{\rm d} = \pi R_{\rm d}^2 \gamma_{\rm GG} = \pi\gamma_{\rm GG} q_{\rm d}^{2} R_{\rm p}^2
\end{equation}
From this relation and the detachment energies calculated by MM simulation in Figure 5, we find $q_{\rm d} \approx 4.0$ (red datapoints and red line in Figure 5a). This is consistent with direct determination of the detachment area as a function of nanoparticle radius, which gives $q_{\rm d} \approx 4.2$ (Figure 5b). By using Eq. (5) elastic energy of one graphene sheet can be obtained as 
\begin{equation}
 E_{\rm el} = \pi R_{\rm d}^2  \frac{\gamma_{\rm GG}}{2} (1-q_{\rm d}^{-2/3}) = \pi R_{\rm p}^2 \gamma_{\rm GG} \frac{q_{\rm d}^2}{2} (1-q_{\rm d}^{-2/3})
\end{equation} 
which, with $q_{\rm d} = 4.0$, gives us the black lines in Figure 5a and 5c. Finally, geometrical similarity requires that, also in case of non cylindrical particles, the effective area of contact between the graphene sheet and the Pt nanoparticle is proportional to the square of the particle radius. This gives us for the Pt-graphene adhesion energy
\begin{equation}
 E_{\rm a} = \pi R_{\rm p}^2 \gamma_{\rm PG} q_{\rm c}^2
\end{equation} 
By matching this with MM simulation data, we find the geometrical factor of $q_{\rm c} = 0.87$ for annealed particles (blue data points and blue line in Figure 5a).  Where $\gamma_{\rm PG} = 0.476$ J/m$^2$ represents Pt-graphene adhesion energy per unit area. For cylindrical particles $q_{\rm c} = 1$. 

\begin{figure*}
	\resizebox{0.9\textwidth}{!}{%
		\includegraphics{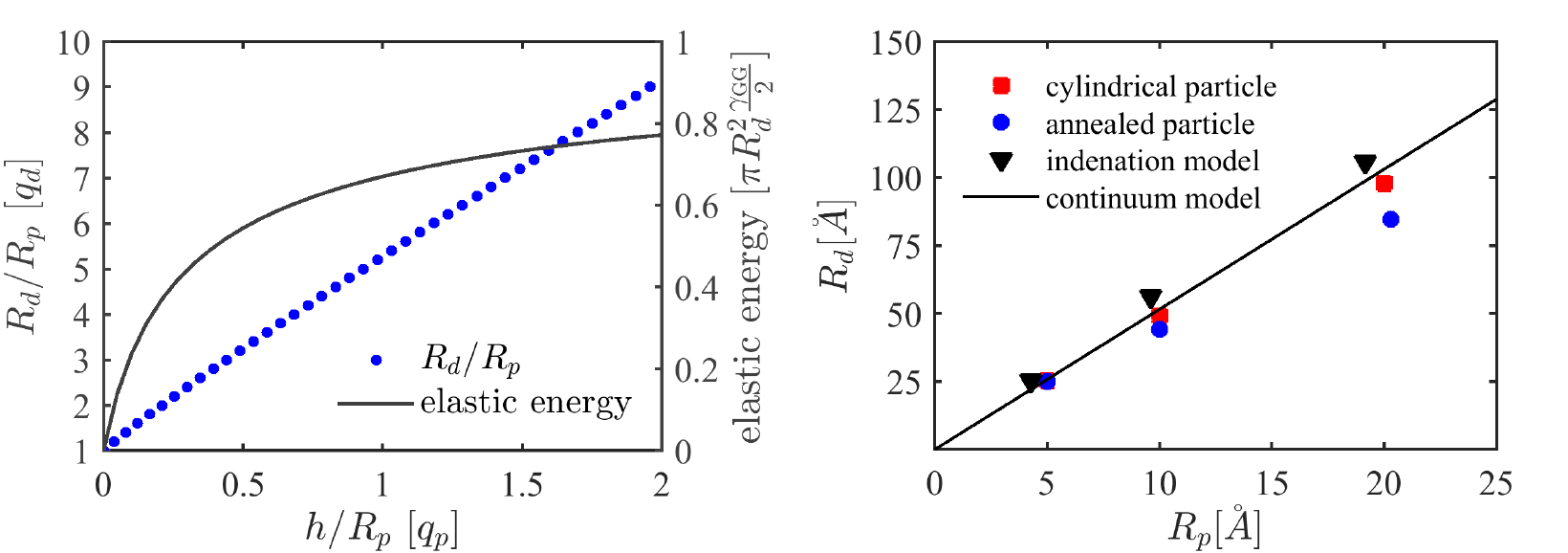}}
	\centering
	\caption{Left: detachment radius and elastic energy of the detached sheet as functions of the 
	particle aspect ratio; right: detachment radius as a function of particle radius; solid symbols:   
	MD simulation data, solid line: continuum prediction for cylindrical particle of aspect ratio $q_{\rm p}= 1$.}
	\label{fig:5}       
\end{figure*}

The elastic energy and the detachment radius are shown in Figure \ref{fig:5} as functions of the nanoparticle aspect ratio. By comparing the continuum results  with the results of our simulations of annealed particles we can show that, for those particles, the elastic energy amounts to about $0.62 E_{\rm d}$, hence $q_{\rm d} \approx 4$ according to the mechanical model. This is in good agreement with the observed proportionality $R_{\rm d} \approx 4.2 R_{\rm p}$ found in the MM simulations in  Figure  \ref{fig:3a}c and Figure \ref{fig:5}. These values correspond to an effective aspect ratio of $q_{\rm p} \approx 0.67$. 

\subsection{Two-particle interaction and mechanical stability of a two-particle system}

As noted in the MD section, if two particles get close and the respective detachment radii overlap, they experience an attractive mutual interaction. A semi-quantitative model can be formulated by approximating the conformation of the graphene sheets around a pair of interacting particles as two detached semicircles of radius $R_d$ connected by a detached ribbon of constant width $w \approx 2R_{\rm d}$ and area $2R_{\rm d} d$ (see Figure  \ref{fig:3b}). The ribbon forms as soon as the particles approach below a critical distance $d_{\rm c}$, and the interaction energy then changes linearly with distance $d$. 

Since the elastic distortion of the graphene sheet over the two detached semicircles is very similar to that around a single particle, we approximate the detachment energy in these regions by that of a single particle. On the other hand, across the detached ribbon the elastic energy is significantly reduced, and we therefore approximate the energy per unit area of the ribbon by $\gamma_{\rm GG}$. The total GG excess energy of the interacting two-particle system is then 
\begin{equation}
E^{\rm 2p}(d,R_{\rm d}) = 1.6 \pi \gamma_{\rm GG} R_{\rm d}^2 + 2 \gamma_{\rm GG} d R_{\rm d}. 
\end{equation}
The detached ribbon forms as soon as this energy falls below the GG excess energy of two non interacting particles, $2 E^{\rm 1p} = 2 \times
1.6 \pi \gamma_{\rm GG} R_{\rm d}^2$. Based on this argument the critical separation follows via $1.6 \pi R_{\rm d} = 2 d_{\rm c}$ as $d_{\rm c} \approx 2.5 R_{\rm d}$ which is in good agreement with the MD data shown in Figure  \ref{fig:3c}. The GG excess energy can then be written as
\begin{equation}
E^{\rm 2p}(d) \approx 2 \gamma_{\rm GG}R_{\rm d}^2 \frac{d_{\rm c}+d}{R_{\rm d}}
\end{equation}
from which the interaction force of the two particles follows as
\begin{equation}
F^{\rm 2p} = \frac{\partial E^{\rm 2p}(d)}{\partial d} \approx 2 \gamma_{\rm GG}R_{\rm d}
\end{equation}
or, expressed in terms of $R_{\rm p} \approx R_{\rm d}/4$
\begin{equation}
F^{\rm 2p} = \frac{\partial E^{\rm 2p}(d)}{\partial d} \approx 8 \gamma_{\rm GG}R_{\rm p}
\end{equation}
The interaction force is, hence, approximately proportional to the Pt nanoparticle radius. Two nanoparticles are stable if this force
is unable to overcome the shear resistance at the Pt-Graphene interface. The shear force required to slide a Pt nanoparticle 
inside the graphene bilayer can be estimated as $F_{\rm c} \approx 2\pi R_{\rm p}^2 \tau_{\rm c}$ where $\tau_{\rm c}$ is the critical shear stress for interface sliding. Equating this to the inter-particle force we find a critical radius below which the particles will slide 
towards each other:
\begin{equation}
R_{\rm p,c} = \frac{4\gamma_{\rm GG}}{\pi \tau_{\rm c}}
\end{equation}
With a typical critical shear stress of $\tau_{\rm c} = 15$ MPa (Figure \ref{fig:2}) we find a critical nanoparticle radius $R_{\rm p} \approx 24$nm. Particles below this radius, if closer than $d_{\rm c} \approx 10 R_{\rm p}$, are likely to spontaneously aggregate. 

\section{Monte Carlo Simulation}
\label{MC}
In the following we study the properties of assemblies of statistically distributed nanoparticles, assuming the nanoparticles are above the critical radius for nanoparticle aggregation and thus on stable locations within the graphene bilayer. Assuming the particles are deposited at random locations with the sole constraint that two particles cannot overlap, we can then use static Monte-Carlo simulation to generate particle assemblies and investigate their geometrical properties. We use this approach to address two questions: (1) what is the minimum concentration of particles that can generate a connected pore space as required for catalytic applications, (2) what is the concentration of particles that yields the smallest overall adhesion of the graphene bi-layer?

In the following, nanoparticles are modeled as disks of radius $R_{\rm p}$ in a two-dimensional continuum, subject to a hard-core repulsion pair potential. Their configurations are obtained using the particle Monte Carlo method, employing the rejection-free Geometric Cluster Algorithm in order to accelerate simulations \cite{liu2004-PRL}. The surface area occupied by a nanoparticle is quantified by counting the number of carbon atoms it covers on the underlying graphene sheet. The same procedure is used to measure detachment areas. For simplicity we assume that the detachment area around a nanoparticle has circular shape with radius $R_{\rm d} = q_d R_{\rm p}$, which provides a lower bound to the actual detachment area in a given multi-particle configuration. Upon increasing the number of nanoparticles, detachment areas form connected components. Component sizes are evaluated employing a standard breadth-first search algorithm on the lattice \cite{moore1959}.

\begin{figure}
	\centering
	\includegraphics[width=0.5\textwidth]{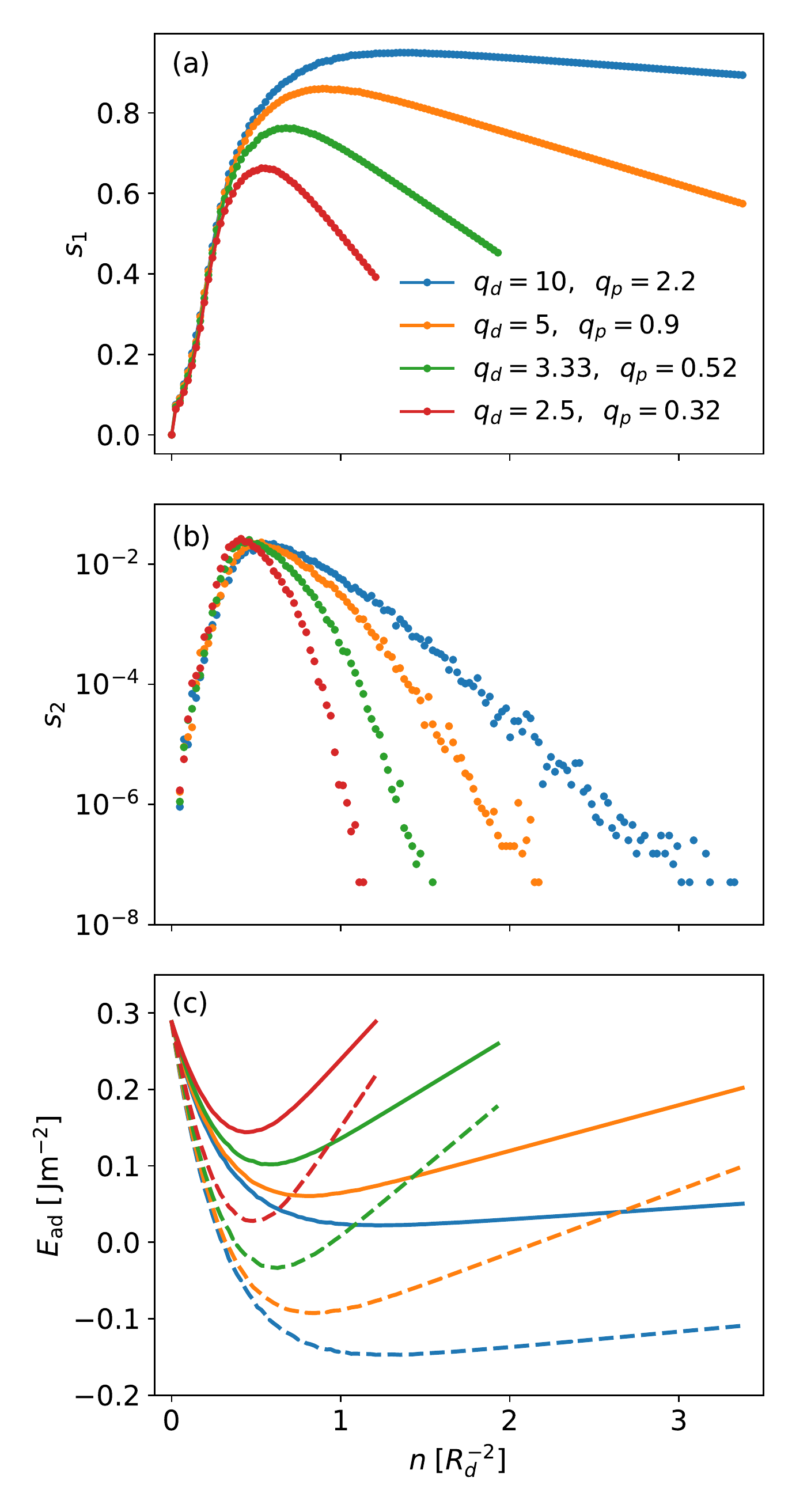}
	\caption{(a) Area fraction of the largest connected pore cluster as a function of particle density, for different particle aspect ratios; (b) Area fraction of the second largest cluster, aspect ratios as in plot (a); (c) Net adhesion energy (energy per unit area required to detach one graphene sheet) as function of particle density, aspect ratios as in plot (a), full line: upper bound given by Eq. (14), top, dashed line: lower bound given by Eq. (14), bottom.}
	\label{fig:6}       
\end{figure}    

We first analyze the question at which nanoparticle density a connected pore space emerges. This is a standard percolation problem. The critical particle density (the percolation threshold) is most conveniently identified by investigating the size $s_1$ of the largest cluster in the system and, more prominently, the size $s_2$ of the {\em second largest} cluster, which is expected to be the largest at the percolation point Figure (\ref{fig:6}). Figure \ref{fig:6} (b) shows that, almost independent from the aspect ratio, percolation occurs when the particle concentration $n$, measured in units of $1/R_{\rm d}^2$ where $R_{\rm d}$ is the detachment radius around a single particle, reaches a value of $n \approx 0.5 R_{\rm d}^{-2}$. The area of the largest connected cluster then continues to increase with increasing particle concentration as remaining areas of direct contact between the two graphene sheets become detached, see Figure \ref{fig:6} (a). The detached area reaches a maximum at a concentration that increases with increasing  ratio of detachment radius to particle radius, but is of the order of $n \approx R_{\rm d}^{-2}$ for all radii. Beyond this maximum, the graphene sheets are completely detached from each other and adding more Pt particles now reduces the available pore space. 
 
Looking at the energy of adhesion required to remove one of the graphene sheets from Pt on one side and separate apart the mutually adhering GG ( going backward from step 3 to step 1 in Figure \ref{fig:3a} ) we can estimate the corresponding energy per unit area as $E_{\rm ad} = f_{\rm GG} \gamma_{\rm GG} + f_{\rm PG} \gamma_{\rm PG} - E_{\rm el}$ where $f_{\rm GG}$ is the area fraction of graphene-graphene contacts, and $f_{\rm PG}$ the area fraction of Pt-graphene contacts, whereas $E_{\rm el}$ is the energy due to elastic distortion of the detached graphene. For the respective adhesion energies we use the values $\gamma_{\rm GG} = 0.288$ J/m$^2$ and, from our DFT calculation, $\gamma_{\rm PG} = 0.476$ J/m$^2$. For the energy of elastic distortion we consider the two limits $f_{\rm PG} << 1$ and $f_{\rm GG} << 1$. In the first case we see mainly isolated nanoparticles. In that case the elastic energy of the detachment area around a nanoparticle is, as shown in Section 2.3, about 0.62 times the detachment energy. The detached area fraction equals $(1 - f_{\rm GG} - f_{\rm PG})$, and the elastic energy is then $E_{\rm el} \approx 0.62 (1 - f_{\rm GG} - f_{\rm PG})\gamma_{\rm GG}$. In the second case, $f_{\rm GG} << 1$, both graphene sheets are no longer in mutual contact and the elastic energy is negligible. This gives us the following expressions for the net energy of adhesion between the two sheets:
\begin{equation}
E_{\rm ad} = \left\{
\begin{array}{ll}
f_{\rm GG} \gamma_{\rm GG} + f_{\rm PG} \gamma_{\rm PG} &,\quad f_{\rm GG} \ll 1\\
f_{\rm GG} \gamma_{\rm GG} + f_{\rm PG} \gamma_{\rm PG} - 0.62 (1 - f_{\rm GG} - f_{\rm PG})\gamma_{\rm GG} &,\quad f_{\rm PG} \ll 1
\end{array}\right.
\end{equation} 

Here, the first expression provides an upper bound and the second expression a lower bound to the actual energy of adhesion. Both bounds are shown in Figure \ref{fig:6} (c) as functions of nanoparticle density. We find a minimum of the adhesion energy which occurs close to the density where the pore space is maximum. The depth of this maximum and hence the maximum reduction in adhesion depend on nanoparticle aspect ratio: pancake-shaped nanoparticles of small aspect ratio provide better adhesion. As a function of nanoparticle density, we find thus three regimes: At low density, we find a non connected pore space concomitant with strong adhesion that decreases with increasing nanoparticle density. At high density, we find a continuous pore space in between the nanoparticles, and adhesion increases with nanoparticle density. At an intermediate density, there is a minimum of adhesion where even spontaneous detachment might occur, in agreement with the experimental finding that decoration with Pt nanoparticles may prevent graphene sheets from aggregating. 

\section{Summary and Conclusions}

We have performed a multiscale simulation study of graphene bi-layers containing embedded Pt nanoparticles. Such composite systems have potential applications for catalysis, but also for the simple purpose of reducing the mutual adhesion of graphene sheets which is important, for instance, if these are to be dispersed as fillers in composites. DFT calculations have provided us with data to parameterize and benchmark potentials for describing the Pt-graphene interface, and to characterize the resistance of this interface to shear displacements of Pt on graphene or vice versa. MD calculations were then performed to determine the energetics and interactions of Pt nanoparticles embedded into graphene, and the results were used to validate a mechanical model which envisages the embedded nanoparticles as rigid cylindrical particles of variable aspect ratio that are inserted between two adherent elastic sheets. With an effective aspect ratio of about 0.67, excellent agreement between the MD results and the predictions of the mechanical model was obtained, both regarding the energetics of the system and the size of the detachment zone surrounding the embedded nanoparticles. The model was then extended to describe the interaction of two nanoparticles
mediated by the intermediate detachment zone, and it was shown that this interaction makes embedded nanoparticles below a critical radius of about $R_{\rm p,c} \approx 24$nm and separation below about 10 $R_{\rm p}$ unstable with respect to spontaneous aggregation. Such aggregation reduces the overall detached area and may push a system with initially connected pore space below the percolation threshold. 

Regarding the geometrical and energetic properties of systems of randomly dispersed nanoparticles above the critical radius, we find that percolation of the detachment areas occurs above a critical nanoparticle density of approximately $1/R_{\rm d}^2$. Above this concentration, the area fraction of the largest connected detachment area first continues to increase, then decreases as the sheets forming the bilayer are completely detached and the addition of further nanoparticles reduces the free space between the graphene sheets. At the same time, the energy of adhesion (the energy needed to separate the bilayer) first decreases, then increases again. In the context of catalytic applications, this allows to increase both the catalytically active nanoparticle surface and the stability of the structure, and trading this off against a reduction in pore space. In the context of using Pt nanoparticles as graphene 'spacers' we find an optimum nanoparticle density where the adhesive energy of the structure is lowest, and thus the separation of the graphene sheets most easy. 

It remains a task for future investigation to extend the present modelling approach from bilayer to multilayer structures. The interaction potentials derived in the present study can serve as a foundation for such a model. However, on larger scales, graphene-mediated interactions between Pt particles located in different layers of a stacked multilayer structure need to be taken into account. A systematic investigation of the resulting complex multi-particle effects and the resulting optimal structures will be a task for future work.  

\section*{Acknowledgments}
SN, KW, MY and MZ acknowledge financial support from DFG under grant no. Za171/11-1. 
M.Z. also acknowledges support by the Chinese State Administration of Foreign Expert 
Affairs under Grant No MS2016XNJT044 for the initial visit to Chengdu 
during which this research was planned. The authors gratefully acknowledge the compute resources and support provided by the Erlangen Regional Computing Center (RRZE).

\section*{Author contributions}
SN and CG performed MD simulations and analyzed MD data using the continuum mechanical model. KW and MY performed DFT simulations. PM performed MC simulations. QL provided unpublished experimental data regarding Pt decorating of graphene which guided the simulations. MZ is responsible for the simulation design. All authors were involved in the preparation of the manuscript. The final version has been read and approved by all authors.

\section*{Conflict of Interest}
the Authors declare no Competing Financial or Non-Financial Interests.

\end{document}